\documentstyle[sprocl,epsfig]{article}

\def\be{\begin{equation}}
\def\ee{\end{equation}}
\def\bea{\begin{eqnarray}}
\def\eea{\end{eqnarray}}

\newcommand\bmat{\left( \begin{array}{cc}}
\newcommand\emat{\end{array}\right)}

\def\msbar{\ifmmode{\overline{\rm MS}} \else{$\overline{\rm MS}$} \fi}
\def\drbar{\ifmmode{\overline{\rm DR}} \else{$\overline{\rm DR}$} \fi}

\def\ti              {\tilde}

\def\D               {\Delta}

\def\s               {\sigma}

\def\sf              {{\ti f}}

\def\tev             {{\rm TeV}}

\begin{document}

\begin{flushright}
  HEPHY-PUB 793/04
\end{flushright}

\vspace*{1.5cm}

\title{Sfermion production at a
Linear Collider at one-loop\\}

\author{Karol Kova\v{r}\'{\i}k}

\address{Institut f\"ur
Hochenergiephysik der \"Osterreichischen Akademie der
Wissenschaften, A-1050 Vienna, Austria}


\maketitle\abstracts{We present the complete one-loop corrections
to the sfermion pair production process $e^+ e^- \rightarrow
\tilde{f}_i\ {\bar{\!\!\tilde{f}}}_{\!j}\;(f = t, b, \tau,
\nu_\tau, u, d, s, c)$ in the Minimal Supersymmetric Standard
Model. Our results also include the previously calculated SUSY-QCD
corrections. We present some details of the renormalization scheme
used. It is found that the weak corrections are of the same
magnitude as the SUSY-QCD corrections at higher energies
($\sqrt{s}\sim 1\tev$). At these energies an important part of the
weak corrections stems from the box contribution. We also include
cross-sections for polarized beams and left-right asymmetry.}

\section{Introduction}
Supersymmetry (SUSY) requires the existence of two scalar
particles (sfermions) $\sf_L$, $\sf_R$ corresponding to the two
chirality states of each fermion $f$. The sfermions of the third
generation play a special role as $\sf_L$ and $\sf_R$ may strongly
mix (proportionally to the fermion mass), forming the two mass
eigenstates $\sf_1$ and $\sf_2$ (with $f= t, b, \tau$) . As a
consequence one eigenstate ($\sf_1$) can have  a much lower mass
than the other one and the mixing angle becomes an important
parameter.
\newline Sfermion pair production in $e^+e^-$ collisions, $e^+e^-
\rightarrow \tilde{f}_i\,\ {\bar{\!\!\tilde{f}}}_{\!j},
(i,j=1,2)$, has been studied extensively phenomenologically.
\cite{exp} The strong interest in sfermion production is mainly
due to the fact that it gives access to one of the fundamental
SUSY breaking parameters $A_f$, the trilinear coupling parameter.
\newline It is clear that in the case of squark production QCD
and SUSY-QCD corrections play an important role.\cite{myletter}
Yukawa corrections \cite{myletter} were shown to be non negligible
either. Here we review the results of the calculation of the full
one-loop corrections to sfermion pair production within the
Minimal Supersymmetric Standard Model (MSSM) as given in
Refs.\cite{myletter,hollik}. A complementary calculation of a
production process with selectrons, smuons and the corresponding
sneutrinos in the final state was also performed in Ref
.\cite{freitas}

\section{Calculation}
The one-loop (renormalized) cross-section $\sigma^{ren}$ is
expressed as 
\begin{equation}
\vspace*{-0.1cm} \s^{\rm ren}(e^+e^- \rightarrow \tilde{f}_i \
{\bar{\!\!\tilde{f}}}_{\!j})=\s^{\rm tree}+\D\s^{\rm
QCD}+\D\s^{\rm EW}\,,
\end{equation}
The electro-weak part can be further split into separate
contributions \footnote{These contributions are not necessarily
gauge invariant. The whole calculation and the separation of the
contributions were carried out in 't Hooft-Feynman gauge.}
\begin{equation}
\vspace*{-0.1cm} \D\s^{\rm EW} = \D\s^{\rm vertex}+\D\s^{\rm
prop}+\D\s^{\rm box}\,,
\end{equation}
where the individual terms denote renormalized vertices, renormalized
propagators and box contributions, respectively.
We restrict the discussion to pure weak corrections which do not
require any inclusion of bremsstrahlung and are thus $\Delta E$
independent. Moreover we can separate the weak part on the basis
of Feynman diagrams in a gauge-invariant way.\newline All the
plots presented here use a scenario based on the SPS1a parameter
input.\cite{spa} As we use the on-shell (OS) values as input, we
transform the \drbar parameters of the SPS1a scenario to obtain
the OS parameters using the relation
\begin{equation}
X^{OS} = X^{\drbar}(Q) - \delta X^{OS}(Q)\,.
\end{equation}

\section{Conclusion}

\begin{figure}[ht]
\begin{picture}(255,105)(0,0)
    \put(14,0){\mbox{\resizebox{5.5cm}{!}
    {\includegraphics{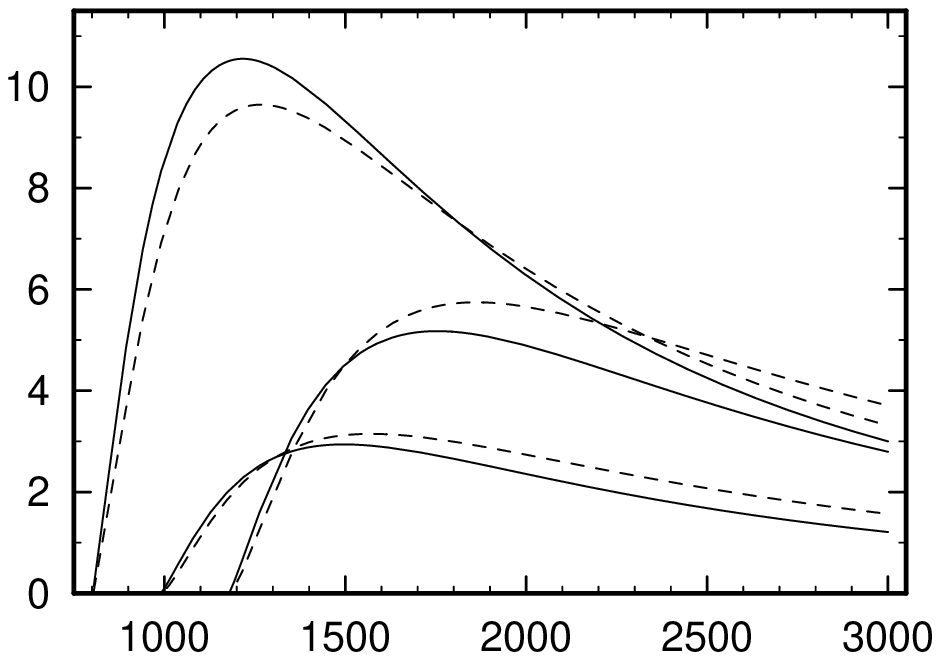}}}}
    \put(116,-5){\makebox(0,0)[r]{$\sqrt{s}~\mbox{{\rm [GeV]}}$}}
    \put(1,14){\rotatebox{90}{$\sigma (e^+ e^-
    \rightarrow \tilde{t}_i\ {\bar{\!\tilde{t}}}_{\!j})~\mbox{{\rm [fb]}}$}}
    \put(106,89){\makebox(0,0)[r]{$\tilde{t}_1\ {\bar{\!\tilde{t}}}_{1}$}}
    \put(116,25){\makebox(0,0)[r]{$\tilde{t}_1\ {\bar{\!\tilde{t}}}_{2}+\,c.c.$}}
    \put(73,60){\makebox(0,0)[r]{$\tilde{t}_2\ {\bar{\!\tilde{t}}}_{2}$}}
    \put(188,0){\mbox{\resizebox{5.5cm}{!}
    {\includegraphics{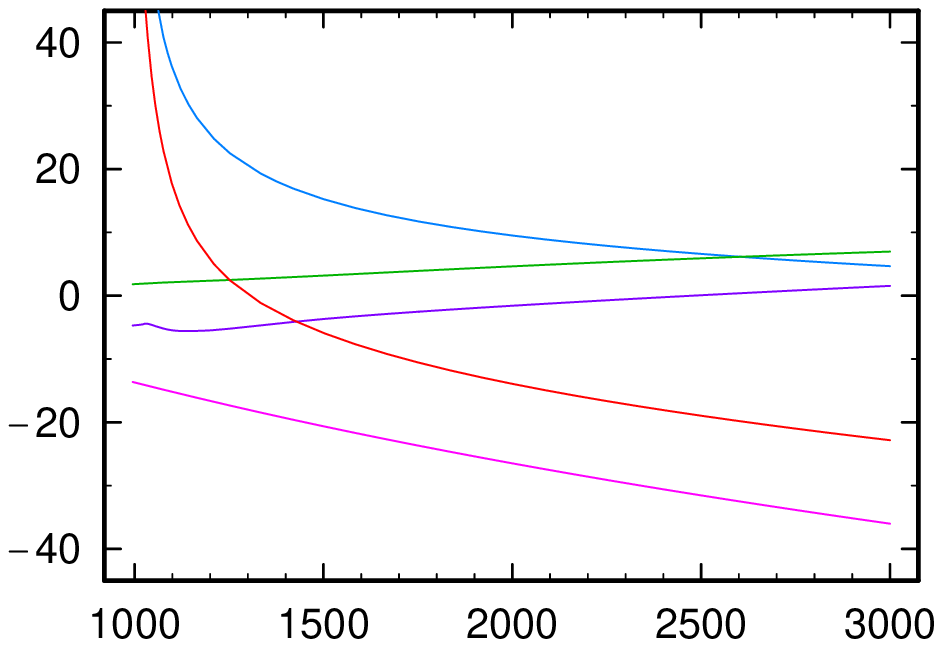}}}}
    \put(300,-5){\makebox(0,0)[r]{$\sqrt{s}~\mbox{{\rm [GeV]}}$}}
    \put(175,22){\rotatebox{90}{$\Delta\sigma/\sigma^0 (\tilde{t}_1\
    {\bar{\!\tilde{t}}}_{2})~\mbox{{\rm [\%]}}$}}
    \put(268,85){\makebox(0,0)[r]{$\Delta {\rm QCD}$}}
    \put(254,28){\makebox(0,0)[r]{$\Delta {\rm box}$}}
    \put(325,48){\makebox(0,0)[r]{$\Delta {\rm total}$}}
\end{picture}
\caption{Stop production with the off-diagonal channel relative
corrections shown.} \label{stops}
\end{figure}

\begin{figure}[ht]
\begin{picture}(255,115)(0,0)
    \put(14,0){\mbox{\resizebox{5.5cm}{!}
    {\includegraphics{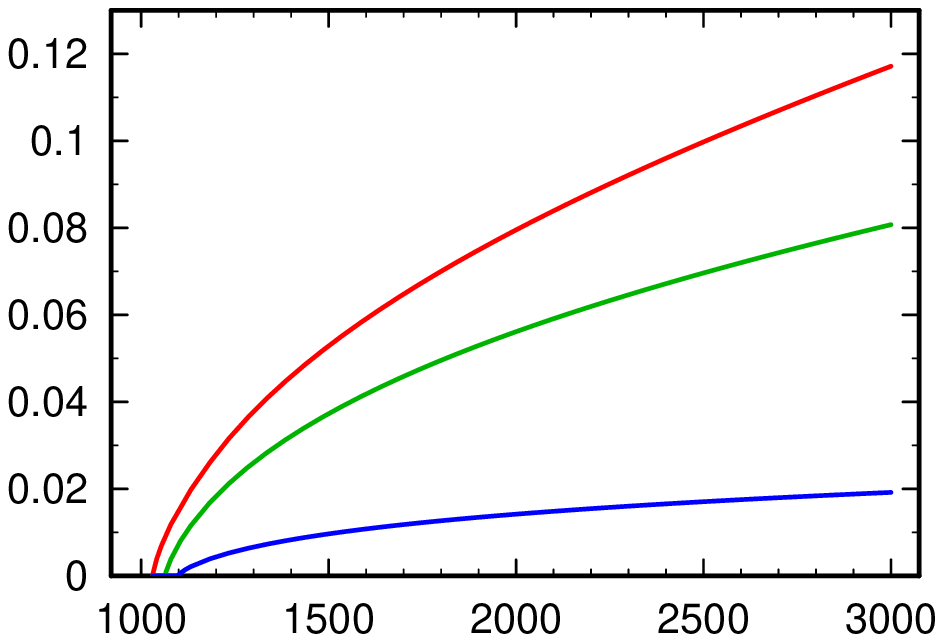}}}}
    \put(116,-5){\makebox(0,0)[r]{$\sqrt{s}~\mbox{{\rm [GeV]}}$}}
    \put(1,50){\rotatebox{90}{$A_{FB}$}}
    \put(125,92){\makebox(0,0)[r]{$\tilde{b}_1\ {\bar{\!\tilde{b}}}_{1}$}}
    \put(125,50){\makebox(0,0)[r]{$\tilde{b}_1\ {\bar{\!\tilde{b}}}_{2}$}}
    \put(152,37){\makebox(0,0)[r]{$\tilde{b}_2\ {\bar{\!\tilde{b}}}_{2}$}}
    \put(188,0){\mbox{\resizebox{5.5cm}{!}
    {\includegraphics{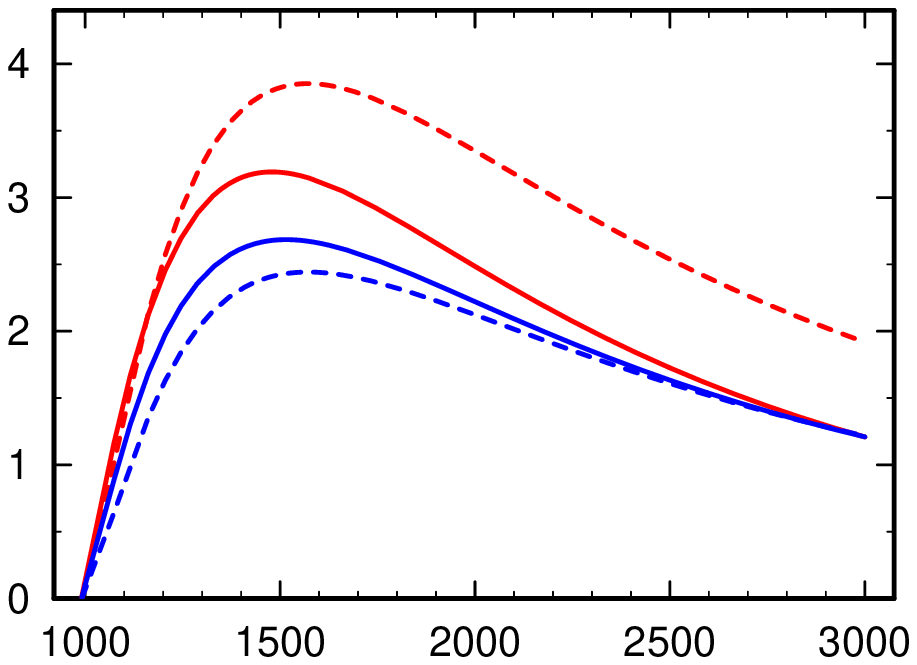}}}}
    \put(300,-5){\makebox(0,0)[r]{$\sqrt{s}~\mbox{{\rm [GeV]}}$}}
    \put(175,22){\rotatebox{90}{$\Delta\sigma/\sigma^0 (\tilde{t}_1\
    {\bar{\!\tilde{t}}}_{2})~\mbox{{\rm [\%]}}$}}
    \put(223,90){\makebox(0,0)[r]{\large $\sigma_L$}}
    \put(244,50){\makebox(0,0)[r]{\large $\sigma_R$}}
\end{picture}
\caption{Left: Forward-backward asymmetry in sbottom production
Right: Polarized cross sections for stop production ($e^-$ beam
polarized)} \label{asypol}
\end{figure}

Electroweak corrections are shown not to be negligible. In
particular at high energy the contibution from box diagrams is
important (see Fig.\ref{stops}). In case of sneutrino production
where there is no QCD contribution and the Yukawa corrections are
not so large (due to $m_\nu$ = 0), the box contribution is the
leading one-loop correction. The production of squarks of the 1st
and 2nd generation is similar because the Yukawa couplings are
zero (we neglect the masses of all quarks of the 1st and 2nd
generation) and so the box diagrams are the largest weak
correction.\newline Some additional information about higher order
corrections can be extracted from polarized cross-sections and
forward-backward asymmetries (see Fig.\ref{asypol}). Here the
forward-backward asymmetry $A_{FB}$ comes only from one loop
corrections as there is no asymmetry at the tree-level.\newline A
further goal is to include also the full QED corrections and thus
complete the full $\mathcal{O}(\alpha)$
calculation.\newline\newline


\end{document}